\documentclass[prd,aps,twocolumn,floatfix,nofootinbib]{revtex4}
\usepackage{graphicx}
\usepackage{dcolumn}
\usepackage{bm}
\usepackage{color}
\usepackage{mathrsfs}
\usepackage{subfigure}

\def\lsim{\mathrel{\mathop
  {\hbox{\lower0.5ex\hbox{$\sim$}\kern-0.8em\lower-0.7ex\hbox{$<$}}}}}
\def\gsim{\mathrel{\mathop
  {\hbox{\lower0.5ex\hbox{$\sim$}\kern-0.8em\lower-0.7ex\hbox{$>$}}}}}

\begin{document}
\newcommand{\mincir}{\raise
-2.truept\hbox{\rlap{\hbox{$\sim$}}\raise5.truept 
\hbox{$<$}\ }}
\newcommand{\magcir}{\raise
-2.truept\hbox{\rlap{\hbox{$\sim$}}\raise5.truept
\hbox{$>$}\ }}
\newcommand{\minmag}{\raise-2.truept\hbox{\rlap{\hbox{$<$}}\raise
6.truept\hbox
{$>$}\ }}

\newcommand{\1}{$\spadesuit$}
\newcommand{\half}{{1\over2}}
\newcommand{\nad}{n_{\rm ad}}
\newcommand{\niso}{n_{\rm iso}}
\newcommand{\ncor}{n_{\rm cor}}
\newcommand{\fiso}{f_{\rm iso}}
\newcommand{\ii}{\'{\'i}}
\newcommand{\bk}{{\bf k}}
\newcommand{\Ocdm}{\Omega_{\rm cdm}}
\newcommand{\ocdm}{\omega_{\rm cdm}}
\newcommand{\OM}{\Omega_{\rm m}}
\newcommand{\OB}{\Omega_{\rm b}}
\newcommand{\oB}{\omega_{\rm b}}
\newcommand{\OX}{\Omega_{\rm X}}
\newcommand{\cltt}{C_l^{\rm TT}}
\newcommand{\clte}{C_l^{\rm TE}}
\newcommand{\calR}{{\cal R}}
\newcommand{\calS}{{\cal S}}
\newcommand{\Rrad}{{\cal R}_{\rm rad}}
\newcommand{\Srad}{{\cal S}_{\rm rad}}
\newcommand{\calPR}{{\cal P}_{\cal R}}
\newcommand{\calPS}{{\cal P}_{\cal S}}
\newcommand{\etal}{{\it et al.~}}
\newcommand{\lya}{{Lyman-$\alpha$~}}
\newcommand{\be}{\begin{equation}}
\newcommand{\ee}{\end{equation}}
\newcommand{\ros}{\rho_{\sigma}}
\newcommand{\planck}{M_{\rm PL}}
\newcommand{\volUnit}{(h/{\rm Gpc})^{-3}}
\input epsf

\preprint{}
\title{Isocurvature, non-gaussianity and the curvaton model}
\author{Mar\'\i a Beltr\'an}
\affiliation{Department of Astronomy and Astrophysics, The University of Chicago, 5460 S. Ellis, Chicago, IL 60637}
\begin{abstract}
Recent analyses of the statistical distribution of the temperature anisotropies in the CMB 
do not rule out the possibility that there is a large non-gaussian contribution to the primordial power spectrum. 
This fact motivates the re-analysis of the 
curvaton scenario, paying special attention to the compatibility of large non-gaussianity of the local type with the current detection 
limits on the isocurvature amplitude in the CMB. 
We find that if the  curvaton mechanism generates a primordial power 
spectrum with an important non-gaussian component, any 
residual isocurvature imprint originated by the curvaton, would have an amplitude too big to be compatible with the current bounds.
This implies that the isocurvature mode should be equal to zero in this scenario and we explore the consequences 
of this inference. In order to prevent the generation of a such a signal, 
the CDM must be created at a late stage, after the curvaton decays completely. 
This is used to constrain the nature of the CDM, 
arriving at a general relation between the temperature of the universe at CDM creation and 
the scale of inflation. It is possible to find 
an absolute maximum for the temperature at CDM creation, which is dependent on the particular inflationary potential. 
For a quadratic potential, we find $T_{\rm cdm}<1.7\times10^{6}$GeV. 
\end{abstract}

\maketitle

\section{Introduction}\label{intro}
An analysis performed by Yadav and Wandelt~\cite{yadav_wandelt}, 
claims the detection of a significant amount of primordial non-gaussianity 
of the local type in the WMAP 3-year data. 
More recently, the WMAP team  \cite{wmap-komatsu} published their latest data release in which the presence of 
a non-gaussian component in the primordial power spectrum is analyzed. 
By using a more sophisticated and conservative mask, the bounds found for the 
non-gaussian contribution are again reconciled with no presence at all, however 
the region of a high non-gaussian contribution has not been ruled out. 
 
 The clear detection of such a signal would automatically rule out the simple single-field  
 inflationary model, since one of its predictions is a gaussian  spectrum for the density fluctuations. 
 We believe that there is not strong evidence against the existence of such a signal and therefore, 
 it is interesting to investigate more sophisticated models  that could explain the origin of 
 the primordial anisotropies along with primordial non-gaussianity at the levels detected.
 
 One suitable candidate, is the curvaton model \cite{curvaton-sloth, curvaton-lyth}. 
 It does predict an almost flat adiabatic spectrum for the density fluctuations along with a 
 possibly large, local, non-gaussian signal and, also possibly,  
 a wholly correlated isocurvature signal which is inside the allowed 
 experimental range up  to now. 
  
 While we wait for the confirmation or rejection of the non-gaussian signal, 
 we study the new constraints imposed in a general curvaton scenario  
 in which the non-gaussian component contributes 
 to the total power spectrum at the level of 
 about $0.1\%$. 
In particular,  we study the compatibility of 
 an eventual detection with the current bounds on primordial isocurvature
 and see what consequences this may bring to the cosmological model.
 
This analysis differs that that carried out in \cite{wmap-komatsu} in the sense that they 
only examine the curvaton as a possible source of isocurvature  whereas we study the possibility that 
this mechanism generate a large non-gaussian signal and maybe, some residual isocurvature.
The approaching angle of this work is more similar to what is presented in \cite{Gordon:2003hw} 
although different assumptions and new data, allow us to draw independent results. 

In section \ref{curvaton} we outline the curvaton model to arrive at an 
expression for the predicted non-gaussianity and isocurvature signals in terms of the 
factor $r$, the ratio of the energy density of the curvaton to the 
total energy density at the decay of the curvaton. 
In section \ref{analysis}, we present the results from several analyses on 
the departures from gaussianity and adiabaticity and find 
which  boundaries are imposed on $r$ by each one of them. 
In section \ref{implications}, we 
combine those constraints to find interesting implications for the decoupling temperature of a cold 
dark matter candidate. 
 Finally, we present our conclusions section \ref{conclusions}.

\section{The curvaton model}\label{curvaton}
The curvaton inflationary model, was proposed as an alternative to the usual mechanism for seeding 
the primordial curvature perturbations \cite{curvaton-sloth,curvaton-lyth}. 
Rather than evolving from quantum fluctuations in the inflaton field, curvature perturbations arise due 
to the presence of an aditional scalar field,  the \textit{curvaton} ($\sigma$). 
This field is practically massless and its energy density, $\ros$, is highly sub-dominant 
at very early times so it is  a spectator during inflation. 
However, after inflation ends and the inflaton decays completely into 
radiation, $\ros$ becomes more and more relevant and the fluctuations in the curvaton field 
emerge as the predominant seed of the observed structure. After that, the Hubble factor 
decreases to a value close to the curvaton mass,  and $\sigma$ starts oscillating. It finally 
decays and the regular Hot Big Bang evolution proceeds. 

There are three general conditions that must be fulfilled in order for the mechanism to work as described 
above, namely:
\begin{enumerate}
\item The curvature perturbation generated by quantum fluctuations in the inflaton field,  $\zeta_{\phi}$ , 
must be negligible compared to the total curvature perturbation.
\item The curvaton cannot trigger a second period of inflationary expansion, therefore, its energy density 
 must be much smaller than the resulting radiation energy density right after inflation.
\item The curvaton is practically massless during inflation  and thus the value of the field is fixed at  $\sigma_*$.
\end{enumerate}
$\zeta_{\phi}$ is the 
curvature perturbation in the uniform density gauge, and it is related to 
the curvature perturbation in any other gauge, $\psi$, by: 
\be\label{zeta0}
\zeta_{\phi}=-\psi-H\frac{\delta \rho_{\phi}}{\dot \rho_{\phi}} 
\ee
which is valid for any generic $\zeta$.

\subsection{Details of the model}
We now describe the basic dynamics of the model adopting 
a parameterization close to that used in the simplest curvaton model \cite{bartolo_liddle}.
We summarize here some of their results that we will be using for our work and 
extend their analysis by using a general form for the inflationary potential and  by 
including the generation of non gaussian perturbations. 
We use the set: $\{V(\phi), m, \Gamma_{\sigma}, \sigma_*\}$, the potential for the inflaton field 
and the mass, the decay rate and the initial value of the 
curvaton field respectively. Later on, we add the non-gaussian contribution and the isocurvature 
fraction parameters, which will be defined in the next sections. 

 The total potential during inflation is:
 \be
V(\phi,\sigma)=V(\phi)+\frac{1}{2}m^2\sigma^2
\ee
which is a good approximation for low values of the curvaton field. 

We adopt the ``curvaton hypothesis", where the inflaton 
curvature perturbation is taken to be less than $1\%$  of the observed value \cite{Dimopoulos:2002kt}. 
Using the COBE normalization at the pivot scale, we can set an 
upper bound for the power spectrum of the inflaton:
\be
\mathcal P_{\zeta_{\phi}}^{1/2}\lesssim 0.01P_{\zeta}^{1/2}\simeq 4.85\times 10^{-7}
\ee
Since \cite{LandL}:
\be
\mathcal P_{\zeta_{\phi}}=\frac{1}{24 \pi^2 M_{\rm{PL}}^4}\frac{V}{\epsilon}
\ee 
(where $ M_{\rm{PL}}=(8\pi G)^{-1/2}\simeq 2.4\times 10^{18}$GeV and $\epsilon\equiv\frac{\planck^2}{2}\left(\frac{V'}{V}\right)^2$), 
we can deduce an upper bound for the energy scale of inflation at the pivot scale.
 Throughout and after inflation, condition 2 implies:
 \be
  m^2\sigma^2\ll V(\phi) \Rightarrow V\simeq V(\phi)
\ee
As long as the inflationary potential is suitable for a Slow Roll hypothesis, we get the upper bound for the value of the 
potential 50 e-folds before inflation ended:
\begin{eqnarray}\label{infmass}
V^{1/4}_{50}&\leq& 6.7\times 10^{15} \rm{GeV},~\rm{or}~\\\label{hmax}
H_{50}&\leq&1.1\times 10^{13}\rm{GeV}
\end{eqnarray}
Since we do not assume any particular shape for the potential, we arrive at these bounds 
by only imposing the slow roll condition for the factor $\epsilon<1$.

Condition 3 implies that the power spectrum for the density 
fluctuations of the curvaton is given by:
\be\label{power}
\mathcal P^{\frac{1}{2}}_{\delta \sigma}(k)\simeq\frac{H_*}{2\pi}
\ee
where $H_*$ is the value of the Hubble factor at horizon exit. 
It is shown in \cite{LUW} that the power spectrum for the density 
contrast for the curvaton field is:
\be\label{power2}
\mathcal P^{\frac{1}{2}}_{\frac{\delta \sigma}{\sigma}}(k)\simeq\frac{H_*}{2\pi \sigma_{\rm exit}}
\ee 
where $\sigma_{\rm exit}$ is the value of the field at horizon exit. Since the field is 
massless during inflation, its value does not evolve with time and we fix $\sigma_{\rm exit}=\sigma_*$ 
for all scales. 
This perturbation spectrum
 is transferred to the total density perturbation by the effects of the 
non adiabatic pressure:
$$\delta P_{\rm nad}=\frac{4\rho_r\ros}{4\ros+3\rho_{r}}(\zeta_{\sigma}-\zeta_r)$$
that arises due to the presence of two different, non interacting, fluids. 
It is shown that the final value of the curvature perturbation inside the 
sudden decay approximation is \cite{LUW}: 
\be\label{zfinal}
\zeta=r \zeta_{\sigma}
\ee
where $r$ is practically the ratio of the curvaton energy density to the total energy density 
at the decay of the curvaton, when $H\sim \Gamma_{\sigma}$. 
In the case where the curvaton dominates completely before its decay, $r=1$.
Using (\ref{zeta0}), (\ref{power2}) and (\ref{zfinal}), we arrive at 
the predicted power spectrum for the curvature perturbation:
\be\label{zfinal2}
\mathcal P_{\zeta}^{\frac{1}{2}}=\frac{2}{3}r\mathcal P^{\frac{1}{2}}_{\delta \sigma/\sigma}\simeq \frac{2}{3}r \left(\frac{H_*}{2\pi\sigma_*}\right)
\ee
which is one of the experimental observables.

One can write the value of $r$ as a function of the parameters of the model. We start with:
\be
r\sim \frac{\ros}{\rho_{\rm r}}\bigg\vert_{\rm dec}=\frac{\ros^{\rm{end}}}{\rho_{\rm r}^{\rm{end}}}\left(\frac{a_{\rm dec}}{a_m}\right)\left(\frac{a_m}{a_{\rm{end}}}\right)^4
\ee
where the sub and super indices  ``end'' refer to the value of the variables at the 
end of inflation and $a_m$ is the scale factor at the time when the 
curvaton starts oscillating, when $H\sim m$. 
Assuming that the decay occurs during a phase of  radiation domination ($r<1$), 
$$
\left(\frac{a(t_1)}{a(t_2)}\right)^4=\frac{H^2(t_1)}{H^2(t_2)}
$$ 
Plugging in the values of the energy densities at the end of 
inflation, we arrive at:
\be\label{r}
r\sim \frac{\sigma^2_*}{6\planck^2}\sqrt{\frac{m}{\Gamma_{\sigma}}}
\ee
Note that the expresion above is independent of the assumed potential for the inflaton. 

In order to prevent a second period of inflation, the curvaton must provide 
a negligible contribution to the energy density, at least until it starts oscillating, at 
$H\sim m$. This implies:
\be\label{cond1}
\frac{1}{3\planck^2}\frac{1}{2}m^2\sigma_*^2\ll m^2 \Rightarrow \sigma^2_*\ll 6 \planck^2
\ee
On the other hand, the decay must occur after the oscillation of the curvaton:
\be\label{decaymass}
\Gamma_{\sigma}<m
\ee
The two factors of equation (\ref{r}) have opposite effects that prevent us from making 
any \textit{a priori} estimation of what the value of $r$ might be.  
Fortunately, $r$ is related to two other observables apart from the power spectrum: 
the amount of non-gaussianity, 
$f_{\rm NL}$ and the isocurvature amplitude, $\fiso$. This will help us put tighter bounds on 
its value. 

\subsection{Non-gaussianity in the curvaton scenario}\label{section-fnl}
The magnitude of the non-gaussian contribution, is parametrized as \cite{Komatsu:2001rj}:
\be\label{nong1}
\Phi=\Phi_L+f_{\rm NL}\Phi_L^2
\ee

where $\Phi$ is the Bardeen potential and is related to the curvature perturbation as:
\be\label{nong2}
\Phi=\frac{3}{5}\zeta=\frac{3r}{5}\zeta_{\sigma}=\frac{r}{5}\frac{\delta \ros}{\ros}
\ee
for super-horizon modes during the matter domination epoch. 
We can expand $\ros$ up to quadratic order:
\be\label{rhos}
\ros=m_{\sigma}^2\sigma^2=m_{\sigma}^2(\sigma_0^2+2\sigma_0\delta\sigma+\delta\sigma^2)
\ee
and plug it into (\ref{nong2}):
\be\label{nong3}
\Phi=-\frac{r}{5}\left(2\left(\frac{\delta\sigma}{\sigma}\right)+\left(\frac{\delta\sigma}{\sigma}\right)^2\right)
\ee
If we identify the first term of (\ref{nong3}) as $\Phi_L$, then, comparing to (\ref{nong1}) we get \cite{LUW}: 
\be\label{fnl}
f_{\rm NL} =\frac{5}{4r}
\ee
 Note that since we have not taken into account the intrinsic, second order 
perturbations in the Bardeen potential, this expression is only valid for high values of $f_{\rm NL}$ which 
dominate over the intrinsic non-gaussianity. This approximation is good enough for our 
purposes since, motivated by the claim in \cite{yadav_wandelt}, we assume study 
the consequences of a local $f_{\rm NL}\sim 100$ (which makes the non-gaussian signal 
 contribute with  $0.1\%$ to the total curvature perturbation). A more 
 precise definition of the non-gaussian contribution can be found in \cite{lyth-rodriguez2005}.

The assumed value for $f_{\rm NL}$ implies a value of $r=0.0125$ which 
also validates the hypothesis assumed in the previous 
section that the curvaton decays during radiation domination.

\subsection{Isocurvature in the curvaton scenario}\label{section-iso}

We focus on the generation of a CDM-isocurvature (CDI) amplitude inside the curvaton 
scenario. 

The CDI mode is defined as:
\be\label{isomode}
\mathcal S_{\rm cdm}\equiv \frac{\delta \rho_{\rm cdm}}{\rho_{\rm cdm}}-\frac{3}{4} \frac{\delta \rho_{\gamma}}{\rho_{\gamma}}
\ee 
This quantity is gauge invariant and, in terms of the individual curvature perturbations (in the flat gauge):
\be\label{isomode2}
\mathcal S_{\rm cdm}=3(\zeta_{\rm cdm}-\zeta_{\gamma})
\ee
We assume that all the species in the universe are radiation when the CDM is created, 
therefore, we can write $\zeta_{\gamma}=\zeta$. 
As in  ref. \cite{LUW}, we take the epoch of creation of the dark matter 
particle as the time after which its number density is conserved. 

There are several parameter configurations that could potentially lead to the 
generation a CDI mode, depending on 
when the CDM is created:

\subsubsection{The CDM is created after the curvaton decays completely}\label{scenario1}
 In this scenario no isocurvature is
generated. The species from which the CDM is hypothetically created, should have thermalized with 
the rest of the universe, carrying the same curvature perturbation, $\zeta$. This is the perturbation that 
will be inherited by the CDM fluid so that $\zeta_{\rm cdm}=\zeta$. Thus, (\ref{isomode2}) is zero. 

\subsubsection{The CDM is created before the curvaton decays and while $f\ll 1$}
 In this scenario, $\zeta_{\rm cdm}=0$ and according to ($\ref{isomode2}$):
\be\label{isomode4}
\mathcal S_{\rm cdm}=-3\zeta
\ee
that is, the isocurvature amplitude is totally anti-correlated to the adiabatic mode, and three times as big. 
This possibility has been ruled out by the analysis of several datasets performed by various groups 
\cite{Amendola:2001ni,iso-dunkley, iso-beltran}. 

\subsubsection{The CDM is created while the curvaton decays} 
This scenario is the most 
model dependent, and we distinguish two possibilities: $i)$ the CDM is created while the curvaton decays and 
$\ros$ is comparable to $\rho_r$. It is predicted that some CDI will be generated, but the precise amount 
depends on the particular creation mechanism. This has been left for further study. $ii)$ the curvaton decays 
{\textit{into}} the CDM. Then, the CDM curvature perturbation will be $\zeta_{\rm cdm}=\zeta_{\sigma}$, and 
using (\ref{zfinal}), we arrive at:
\be\label{isomode3}
\mathcal S_{\rm cdm}=3\left(\frac{1-r}{r}\right)\zeta
\ee
Which leads to an isocurvature mode that is correlated to the adiabatic one, has the same scale dependence, 
and has an amplitude that depends on the ratio of curvaton to radiation energy density. 
The isocurvature amplitude is small if  the curvaton is close to dominating the 
energy density of the universe when it decays. If the curvaton decays while it is still sub-dominant, the resulting 
entropy mode could be large. 

In section \ref{section-fnl} we mentioned the value of $r$ corresponding to a $0.1\%$ non-gaussian contribution in the 
primordial power spectrum: $r=0.0125$. In what follows, we see that the current upper bounds on a curvaton-like 
isocurvature signal are in tension with such a value, that indicates that the curvaton energy density is not 
close to domination when it decays. This starts pointing towards our conclusion that 
{\textit{if the curvaton mechanism is responsible for a large non-gaussianity}}, then, the same mechanism 
cannot generate any isocurvature signal whatsoever. We see this in more detail in the next section. 

\section{Data Analysis}\label{analysis}
In this section we describe the existent experimental constraints on $f_{\rm NL}$ and $\vert \frac{\mathcal S_{\rm cdm}}{\zeta}\vert$. 
The most restrictive results to the date are those reported by the WMAP team in their 5-year report, which we quote along with 
significant analyses performed by other groups. 
\subsection{Non-gaussianity}
In \cite{yadav_wandelt}, Yadav and Wandelt analysed the two main science channels of  the WMAP 3rd-year data up to $\ell_{\rm max}=750$. 
Using the estimator described in \cite{yadav-estimator} and the {\textit{Kp0}} foreground mask, they claimed a detection of a non-gaussian, local,  component of:
\be\label{fnl-bounds}
26.91<f_{\rm NL}<146.71~\rm( at~95\%~c.l.)
\ee
More recently,  with a better understanding of the point source contamination and a more restrictive 
\textit{KQ75} mask \cite{Gold-wmap}, the WMAP team reported an allowed region \cite{wmap-komatsu}:
\be\label{fnl-bounds-wmap}
-9<f_{\rm NL}<111~\rm( at~95\%~c.l.)
\ee
from the analysis of the bispectrum and using the same estimator as Yadav and Wandelt. The 
maximum multipole number used in \cite{wmap-komatsu} is $\ell_{\rm max}=700$. 
The WMAP team also reports a second result from the analysis of the Minkowski functionals following the 
method described in \cite{minkowski}.  With this method, they find a new result which is in slight tension 
with the previous one:
\be\label{fnl-bounds-wmap-mink}
-178<f_{\rm NL}<64~\rm( at~95\%~c.l.)
\ee
Recently, Hikage et. al also studied the non-gaussianity in the WMAP 3rd-year data with 
the Minkowski functionals  \cite{Hikage:2008gy} to find:
\be\label{fnl-bounds-hikage-mink}
-70<f_{\rm NL}<91~\rm( at~95\%~c.l.)
\ee
which seems in agreement with the WMAP team analysis with the Minkowski functionals. 

We show all these different results to highlight two points: one is the fact that there is al lot left to understand 
about primordial non-gaussianity and the statistical methods to detect it. The other is that none of 
the results quoted above rule out a possibly large (of order $0.1\%$ ) non gaussian component 
and that is the reason why we believe it is worth exploring the curvaton model as 
the responsible for a large $f_{\rm NL}$.

Note that a part of the two sigma allowed region in (\ref{fnl-bounds-wmap}) cannot be explained in terms of the 
curvaton dynamics because $r$ has a natural upper bound of 1. This value of $r$ corresponds 
to $f^{\rm min}_{\rm NL}=-1.25$ (taking into account the second order corrections 
to $f_{\rm NL}$ described in \cite{lyth-rodriguez2005}).

\subsection{Isocurvature}\label{isocurvature}

With two more years of data gathering  the WMAP team achieved stringent limits on the 
parameters of the basic cosmological model and some extensions such as departure from 
adiabaticity. They use the parameter $\alpha$ to measure the amplitude of the isocurvature mode in 
the scenario in which the curvaton decays into dark matter generating a mode with amplitude 
described by (\ref{isomode3}). 
The parameter $\alpha$ is defined as \cite{alphadef}:
\be
\alpha\equiv\frac{f_{\rm iso}^2}{1+f_{\rm iso}^2} 
\ee
where $\fiso$ is the ratio the isocurvature and adiabatic amplitudes at the 
pivot scale. In the notation defined in the sections above, $\fiso=\vert \frac{\mathcal S_{\rm cdm}}{\zeta}\vert$. 
This parametrization has the advantage that $\alpha$ is naturally bounded to take 
a value $\in [0,1]$. Some controversy exists about whether or not some isocurvature 
models are artificially favored by this choice \cite{Trotta:2005ar, iso-beltran} but our model 
is already selected, so this should not be an issue for the analysis. 

The WMAP team finds the following upper bound for $\alpha$ inside the frame of 
the curvaton model (totally correlated \footnote{In \cite{wmap-komatsu}, they 
use the ``anti-correlated" result, but this is due to the sign convention used for the 
definition of the correlation amplitude.}, same scale dependence of the adiabatic and 
isocurvature amplitudes):
\be
\alpha < 0.0037 ~\rm( at~95\%~c.l.)
\ee
using their 5-year CMB data plus Type Ia supernovae \cite{supernovae} and 
Baryon Acoustic Oscillation data \cite{bao}.
This bound improves previous results obtained with for the 3rd-year WMAP data 
release plus other data, including large scale structure \cite{iso-dunkley} by a factor 2.5. 
However, the tightest bound to the date is found when the Lyman-$\alpha$ forest data is 
included, $\alpha < 0.0015 ~\rm( at~95\%~c.l.)$ \cite{iso-beltran}.

In any case, the difference in the derived value of the ratio of energy densities is minimal. 
Taking into account (\ref{isomode3}), we see:
\be
r=\frac{1}{1+\frac{\fiso}{3}}\simeq 1-\frac{\fiso}{3}\simeq 1-\frac{\sqrt \alpha}{3}
\ee
Thus, we finally arrive at the experimental result that if the curvaton decays into dark matter, 
it must do so when its energy density is just about to become the dominant component in the 
universe, \textit{i.e.}, the value of $r$ must be bounded by:
\be
0.98< r<1
\ee
(using the WMAP bound on $\alpha$). This bound corresponds to a non-gaussianity contribution of:
$-1.21>f_{\rm NL}>-1.25$ which is clearly inside the 2 $\sigma$ region derived by \cite{wmap-komatsu}. 
However, it is one sigma away from the central value quoted in Table 5 of \cite{wmap-komatsu}. 
Although this 
fact is not significant from a statistical point of view, 
it suggests the intriguing possibility that the non-gaussianity values could be in strong conflict 
with the value of $f_{\rm NL}$ derived from the curvaton-like isocurvature bounds. 

In what follows, we focus in this potential scenario, and study the consequences of 
an eventual detection of a high non-gaussian component inside the curvaton framework. 

\section{Implications for CDM}\label{implications}

As described in the previous section, the detection of a large  non-gaussianity 
would restrict the allowed bound for the ratio $r$ to a region incompatible with 
the range suggested by current limits on the isocurvature amplitude.  

Thus, if one accepts that the curvaton is the model responsible for the generation of the primordial non gaussianity, 
one must accept as well the fact that the curvaton mechanism proceeds according to subsection \ref{scenario1} and 
no isocurvature imprint is left in the CMB. 
This leads to two important consequences:
\begin{enumerate}
\item The CDM is not the direct decay product of the curvaton.
\item The CDM must be created after the curvaton field has decayed.
\end{enumerate}
The last point implies a severe constraint on the possible values of the temperature of the universe 
at the epoch of creation of dark matter. For thermal relics, this translates directly onto a constraint on their 
mass. Inversely, the detection of a particular dark matter species, would give us some hints about the 
precise dynamics of the curvaton mechanism. 

Firstly, let us point out that consequence 2 implies:
\be
H_{\rm cdm}< \Gamma_{\sigma}
\ee
where $H_{\rm cdm}\simeq T^2_{\rm cdm}/\planck$ is the Hubble factor at the epoch of DM creation, at a 
temperature $T_{\rm cdm}$. 
The upper bound for $T_{\rm cdm}$ is then, 
\be
T_{\rm cdm}<\sqrt{\Gamma_{\sigma}\planck}
\ee
Thus, a constraint in $\Gamma_{\sigma}$ could have strong implications for the particular mechanism of dark matter
generation. 

Note that the possibility that the CDM was created much before the curvaton decayed, had been ruled out already, 
as we mention in section \ref{section-iso}. However, this fact alone was not enough to impose severe constraints on $T_{\rm cdm}$. 
In the first place, one could always appeal to the attractive possibility that the curvaton decayed indeed into the dark matter- an instance that 
has been ruled out under our assumptions. Secondly, without the additional constraint provided by 
the assumption of large non-gaussianity, it is not possible to derive the tight boundaries on the decay epoch of the curvaton 
that we describe below. Therefore, it is this particular combination of extensions to the standard inflationary scenario that 
allows us to draw the conclusions that we present as our main result. 

In what follows, we go through further considerations about the model that help us reduce the dimension 
of the space of parameters and arrive at a simple relation between the maximum $T_{\rm cdm}$ and 
the scale of inflation. 

We can solve for $\Gamma_{\sigma}$ in (\ref{r}) and write $r$ as a function of $f_{\rm NL}$:
\be\label{gammasigma}
\Gamma_{\sigma}=m\left(\frac{5f_{\rm NL}}{4}\frac{\sigma^2_*}{6\planck^2}\right)^2
\ee
Its value depends on $m$ and $\sigma_*$ which are in principle very loosely bounded. 
Below we will show how two additional considerations relate the value of two of these parameters to 
each other.  

\subsection{The value of $\sigma_*$}
Even though the bound (\ref{cond1}) trims off a large range of possible initial values of the curvaton field, 
it is not informative enough for the purpose of studying the limits of the mass of a cold dark matter candidate. 
Luckily, it is possible to find a lower bound for  $\sigma_*$ if we consider the 
effect of the de Sitter vacuum on the field. 
It is shown in \cite{Linde:1982uu} that 
the value of the variance of an almost massless field after a period of 
exponential expansion approaches the limit:
\be\label{variance}
\langle \sigma^2\rangle=\frac{3H^4}{m^2 8 \pi^2}
\ee
if the expansion has occurred for a long enough time (i.e., the number of e-folds before $N\simeq 50$ is large).
It is sensible then, to expect that the field at least has moved that much away from its minimum, and thus,  
this value can be taken as the lower bound for $\sigma_*$ after inflation:
\begin{equation}\nonumber
\frac{3H^4}{m^2 8 \pi^2}<\sigma^2_*<6\planck^2 \label{limit-mass}
\end{equation}
As we will see below, this inequality leads to a lower bound on the mass of the curvaton which 
noticeably shrinks the allowed region in the space of parameters. 

If we do not assume that inflation lasted for many more e-folds than those we have causal access to, then the limit above 
weakens. The inequality:
\be
\langle \sigma^2\rangle<\sigma^2_*<6\planck^2
\ee
does not constrain the value of $\sigma_*$ strongly and the allowed range for the mass of the curvaton is big. 
We are left with the single constraint $m>\Gamma_{\sigma}$ which this leads to 
a much smaller lower bound for the mass. We will see that this scenario is somewhat less interesting 
than the one in which $m$ is more restricted because the former calls for a certain amount of 
fine tuning if CDM is required not to leave any isocurvature trace. 
Thus, we treat this ``short inflation scenario" as a caveat that we discuss at the end of this section, while we assume 
the long inflation scenario in what follows. 

\subsection{Constraints coming from the power spectrum}
According to our assumptions, the curvaton must be responsible for most of the 
power at the pivot scale. 
Using (\ref{zfinal2}) and the relation of $r$ to $f_{\rm NL}$, we can find an additional constraint on the parameters that determine the 
density contrast. Equation 
(\ref{zfinal2}) becomes:
\be
\mathcal P^{1/2}_{\zeta}=4.8\times 10^{-5}=\frac{5}{12\pi f_{\rm NL}}\frac{H_*}{\sigma_*}
\ee
Solving for $\sigma_*$  we get:
\be
\sigma_*=\frac{5}{12\pi f_{\rm NL}}\frac{H_*}{4.8\times 10^{-5}}=27.6 H_*\cdot \left(\frac{100}{f_{\rm NL}}\right)
\ee
The lower bound on $\sigma_*$, (\ref{variance}), provides a powerful relation between 
the energy scale for inflation and the mass of the curvaton:
\be\label{mass-range}
7\times10^{-5}f_{\rm NL}<\frac{m}{H_*}<1
\ee
where the upper bound comes from the requirement that the curvaton starts oscillating after the pivot scale 
 exits the horizon. As a matter of fact, this bound is even tighter because the curvaton should not oscillate before 
 the end of inflation, and $H_{end}<H_*$. However, the precise relation between these two values of the Hubble 
 factor depends on the particular inflationary potential. Since we wish to keep the analysis valid for a generic potential, 
 we simply use (\ref{mass-range}) keeping in mind the previous statement. 

We also considered bounds coming from the tilt of the spectrum. Even though it is not explicitly quoted in \cite{wmap-komatsu}, 
looking at the two dimensional probability distribution of $\alpha$ and $n_s$ in Fig. 9 of that paper, we estimate a marginalized 2 sigma 
region of: $0.93\lesssim n_s\lesssim1$. 
The spectral index in this model is related to the primordial parameters by \cite{curvaton-lyth}:
\be
n_s= 1+2\frac{V_{\sigma \sigma}}{3H_*^2}-2\frac{\dot H_*}{H_*^2}
\ee
We must make sure that the two additional factors in the RHS of this equation are not in conflict with the experimental bound. 

The second factor, is directly related to  the ratio of primordial tensor to scalar amplitude \cite{lyth-tensor}.
Due to the low energy during inflation, the curvaton model predicts non observable 
primordial tensor modes, which implies $\tau<0.07$ \cite{lyth-tensor}. Thus, inside this workframe, it is expected that the ratio is limited from above by 
 $2\frac{\dot H_*}{H_*^2}<0.01$ which would be irrelevant given the current precision in the determination of the spectral tilt. 

On the other hand, in order to prevent the generation of a blue tilt in conflict with observations, we impose the mild 
requirement that:
\be
2\frac{V_{\sigma \sigma}}{3H_*^2}<\mathcal O (10^{-2})\Rightarrow \frac{m}{H_*}<0.1
\ee
which reduces the allowed range (\ref{mass-range}) (this was also pointed out in \cite{curvaton-lyth}). 

Plugging in our latest relations into the decay rate, we get:
\be\label{gammalast}
\Gamma_{\sigma}=b\cdot H_*^5\left(\frac{1.59\times 10^6 }{\planck^2f_{\rm NL}}\right)^2
\ee
where $b$ is a number that parametrizes the  curvaton mass, $m=b\cdot H_*$, and can take any value 
inside the interval $(7\times10^{-5}f_{\rm NL} ,0.1)$.

 After all the considerations, we can see that $\Gamma_{\sigma}$ depends mainly 
on the scale of inflation, and only mildly on the precise value of the mass of the curvaton. Indeed, 
the corresponding decay temperature, only depends on $\sqrt{b}$.

\subsection{Mass of the CDM candidate particle}

We stick to the convention described in section \ref{section-iso} regarding the definition of the 
time of CDM creation. For candidates such as WIMPS, their creation corresponds to their freezing-out 
of thermal equilibrium in the early universe. In this case, 
we use the relation \cite{jungman-rept}:
\be\label{tcdm}
T_{\rm cdm}\simeq \frac{m_{\rm{cdm}}}{20}\Rightarrow m_{\rm{cdm}}<20\sqrt{\Gamma_{\sigma}\planck}
\ee
to find an upper bound to the {\textit{mass}} of the CDM candidate. Note that the mass of the heaviest thermal 
relic corresponds to $m_{\rm cdm}\simeq 240$ TeV \cite{griest-kamionkowski}. 
Heavier dark matter species must have been created out of equilibrium, in order to not over close the universe. 
In this case, the relation of the mass to the creation temperature is not (\ref{tcdm}), and the limits apply to 
the temperature of the universe at the era of their creation. 

For simplicity, during this section we will assume that the non-gaussian contribution to the primordial 
power spectrum is $f_{\rm NL}=100$. The results are easily generalizable to a different value 
of $f_{NL}$, as long as it fulfills the large-non-gaussianity requirement specified in section \ref{analysis}. 
Also, we show below that the results are pretty robust against variations of this parameter. 
In order to minimize the size of the space of 
parameters, we fix the value of $b=0.1$ and check at the end that the effect of this factor on the results is 
negligible. Note that in any case, this fixed value of $b$ will generate the most conservative results as it sets the 
highest possible upper bound for $\Gamma_{\sigma}$. 

In order to prevent a neutrino isocurvature mode \cite{LUW}, the decay of the curvaton must 
occur before the neutrino decoupling era, at $T_{\nu}\simeq 1$MeV:
\be
\Gamma_{\sigma}>\frac{T_{\nu}^2}{\planck}\Rightarrow H_*>1.4\times 10^{8}\rm{GeV}
\ee 
which leaves a narrow margin for the energy scale of inflation:
\be
6.7\times10^{16}\rm{GeV}> V^{1/4}>2.4\times 10^{13}\rm{GeV}
\ee 
as was pointed out in \cite{Huang} (note that this range could be shrunk further if $\epsilon \ll 1$, but 
this depends on the particular inflationary potential). Combining equations (\ref{tcdm}) and (\ref{gammalast}), 
we arrive at a new relation that connects inflationary phenomenology and 
 particle physics beyond the standard model:
 \be
m_{\rm cdm}<3.2\times 10^5 \planck^{-\frac{3}{2}} H_*^{\frac{5}{2}}
\ee
It applies for masses smaller than $\sim 10^5$ GeV, that is, for $H_*<9.6 \times 10^{10}$ GeV. 
For higher values of the Hubble factor, the bound is imposed on $T_{\rm cdm}$:
 \be
T_{\rm cdm}<1.6\times 10^4 \planck^{-\frac{3}{2}} H_*^{\frac{5}{2}}
\ee
with an absolute maximum for $T_{\rm cdm}$ is  $T_{\rm cdm}< 1.8 \times 10^8$ GeV. 

\begin{figure}[h!]            
\begin{center}
\includegraphics[angle=0, width=0.45\textwidth, height=0.25\textheight]{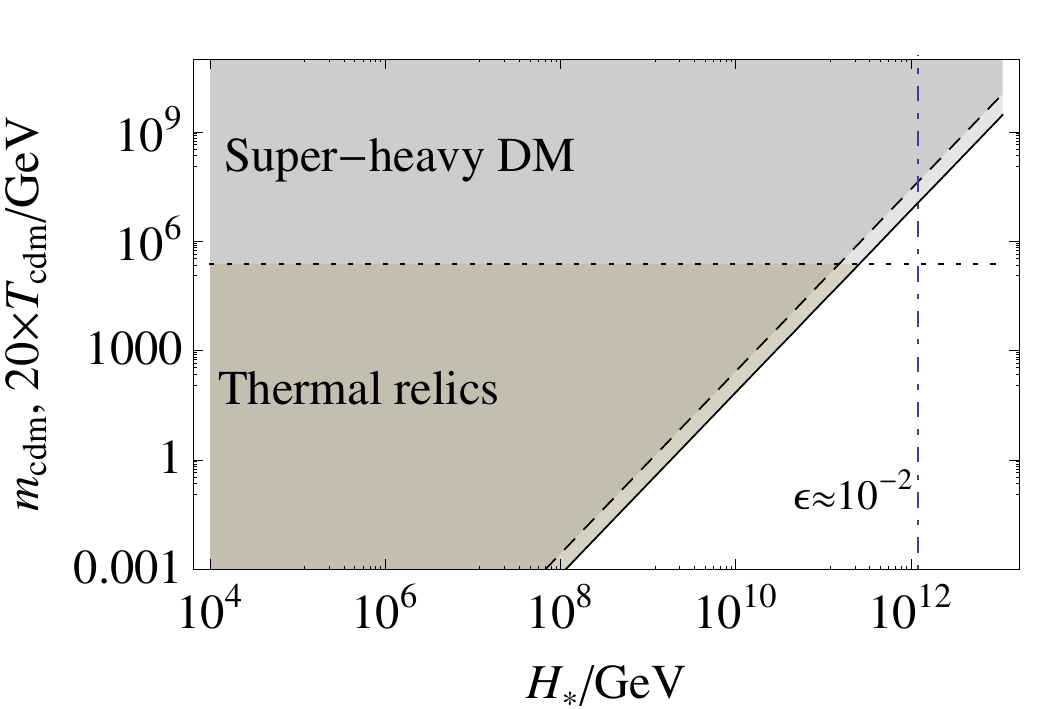}
\end{center}
\caption{Upper bound on the mass of a CDM candidate as a function of the inflationary scale. We used 
the two bounding values for the mass of the curvaton: $m^{max}=0.1H_*$ (dashed), and $m^{min}=7\cdot 10^{-3}H_*$ (solid). 
The dotted line corresponds to $m_{\rm cdm}=2.4\times 10^5$GeV, the value above which the bounds apply to the temperature of DM creation. 
The vertical dot-dashed line shows the tighter constraint on the scale of inflation coming from assuming a quadratic potential for the inflaton. 
The value of  $f_{\rm NL}$ has been fixed to $100$.}
  \label{mass}
\end{figure}

 In Fig. \ref{mass}, we plot these bounds as a function of the inflationary scale, $H_{*}$. 
 The shaded areas above the solid line correspond to values of the 
 parameters that would generate an isocurvature mode, for a curvaton mass of $m=0.1\times H_*$.  
The dotted line limits the region in which the creation temperature of CDM 
can be directly related to its mass via equation (\ref{tcdm}). Above this line we can only constrain the temperature 
of the universe at the epoch of CDM creation. 
To assess the generality of the model, we also plot 
the limit resulting for the minimum possible mass of the inflaton, corresponding to $b=7\times10^{-3}$ 
and see that the conclusions are practically the same (dashed line).

Note that the consequences of the eventual detection of a CDM particle of a mass of around $100$ GeV, would contract  the allowed range 
by two orders of magnitude, leaving a narrow range of $\sim 4\cdot 10^9-10^{13}$ GeV for the possible energy scale for inflation. 
It is also remarkable that once one assumes a particular shape for the inflaton potential, the parameter $\epsilon$ can be determined. Then, 
since its value is generally smaller than 1, the upper bound for the scale of inflation will lower, shrinking further the allowed region. 
In particular, for a quadratic potential, we would find $\epsilon\simeq 10^{-2}$, and $H_{50}$ (or the maximum value for $H_*$) would be one order 
of magnitude smaller. We show this with a dot-dashed vertical line in 
Fig. \ref{mass} . In this case, the absolute upper limit for the creation temperature of a CDM candidate is, 
$T_{\rm cdm}< 1.7 \times 10^6$ GeV. 

The two temperature upper bounds would be relevant for primordial black hole (PHB) production, although 
the evolution and final density of such a species is highly dependent on the particular details of each model. 
Furthermore, small size, quickly evaporating PBH's could be generated at very early times leading to a large entropy release 
(see, e. g. \cite{bellido-pbhs}) that 
would modify the conclusions of this analysis. Therefore the possibility of PBH generation and 
the compatibility with the model, is left for future work. 

Gravitationally generated wimpzillas are created mostly at the end of inflation, for inflationary scales of $H_{end}\sim 10^{13}$ GeV  \cite{wimpzillas}. 
This corresponds to a temperature above the upper bound, and thus they are incompatible with this curvaton model. 
Nevertheless, wimpzillas can also be generated during reheating and preheating leading to model dependent masses or 
temperatures of creation. The existence of these second kind of wimpzillas would not be in conflict, in general, with 
the curvaton bounds. 

{\textit{``Short inflation" scenario}}: If inequalities (\ref{limit-mass}) no longer apply, the lower bound on $m$ comes from demanding: 
\be
\Gamma_{\sigma}=m\cdot H_*^4\left(\frac{1.59\times 10^6 }{\planck^2f_{\rm NL}}\right)^2> T^2_{\nu}/\planck
\ee
which becomes the most extreme for the maximum value of $H_*$, in eq. (\ref{hmax}). 
In that case, the above inequality is fulfilled as long as:
\be
m>1.14\times 10^{-28}\planck
\ee
which corresponds to a value of $b\sim2\times 10^{-22}$. 
Essentialy, extending the allowed range for $b$ opens up the space of parameters and the relation between 
$T_{\rm cdm}$ and the scale of inflation is incomplete without the specification of the value of $b$. 
However, it is significant that the range opens up towards its lower end since 
allowing $b$ to take  smaller values only narrows 
down the allowed region of Fig. \ref{mass}. This leaves less and less time for the CDM to freeze out 
after the decay has occurred and would lead ultimately to a CDM that has to decouple just before neutrinos do. 
This makes the model fine tuned as, to some extend, less likely because the masses of the curvaton and the CDM 
particle which are unrelated a priori, must conspire to prevent the generation of isocurvature modes.  

\vspace{3mm}

\begin{figure}[h!]
\begin{center}
\includegraphics[angle=0, width=0.45\textwidth, height=0.25\textheight]{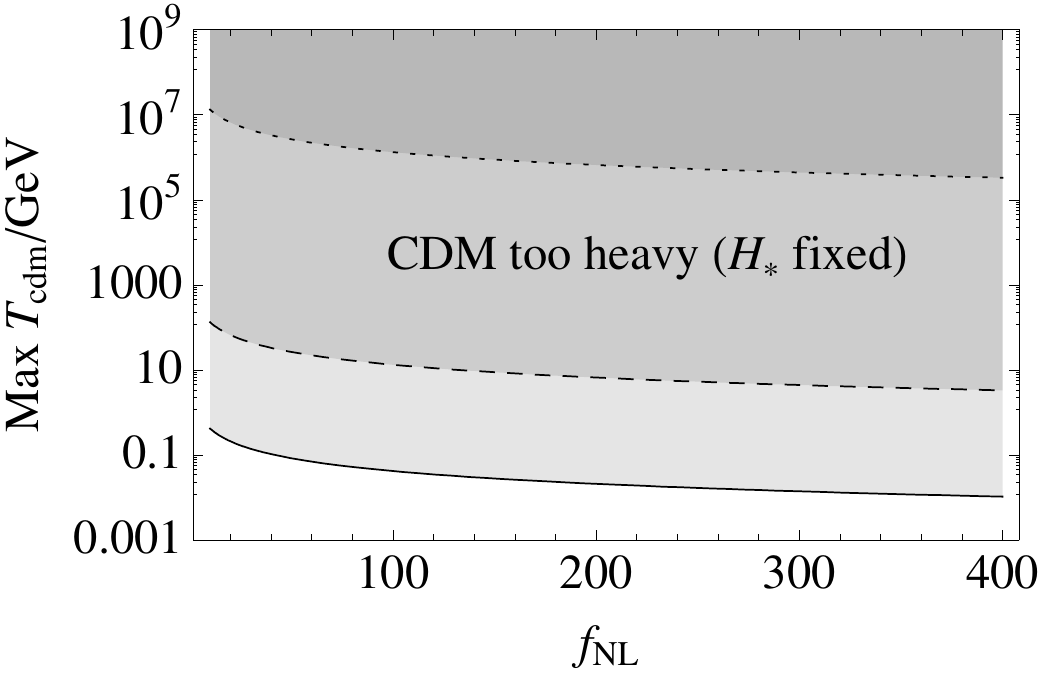}
\end{center}
\caption{Upper bound on the mass of a CDM candidate as a function of non-gaussian fraction. We used three different 
fixed values for the scale of inflation: $10^9$ GeV (solid), $10^{10}$ GeV (dashed), $10^{12}$ GeV (dotted). The value of 
$b$ used is $b=0.1$}
  \label{mdef}
\end{figure}

Now we study how the results change when we drop the assumption that $f_{\rm NL}= 100$. In Fig. \ref{mdef}, 
we plot the upper bound on  $m_{\rm cdm}$ as a function of $f_{\rm NL}$ for several fixed values of the energy scale. 
The range used for the possible value of the non-gaussianity is:
$$
10<f_{\rm NL}<400
$$
where the upper bound is an approximation of the result in \cite{Huang}, where they set the limit $f_{\rm NL}<522 \cdot \tau^{1/4}$ 
 for the curvaton model of inflation. 
We see how, for a fixed inflationary scale, the maximum allowed $T_{\rm cdm}$ hardly varies with $f_{\rm NL}$. 
This is again due to the strong dependence of  $T_{\rm cdm}$  on the scale of inflation, while it only depends 
linearly on the non-gaussianity factor.  With this we show that the analysis is robust against variation of the non-gaussianity 
contribution. This is particularly convenient, given the existing uncertainties in the estimation of $f_{\rm NL}$.

 \section{Conclusions}\label{conclusions}

Given the recent advances on the detection of primordial non-gaussianity in the CMB, 
we examined the implications of an eventual observation of a large non-gaussian signal 
($f_{\rm NL}\sim 100$) for a very simple curvaton model. 
In particular, we studied the residual isocurvature signal that could be imprinted in this scenario. 

It is found that, if the curvaton is indeed the mechanism responsible for seeding a primordial
spectrum with an important non-gaussian contribution, 
it cannot induce as well any trace of primordial isocurvature.  
This leads to two important conclusions that help confining the model: 1. The curvaton 
field cannot decay into dark matter, 2. The dark matter species must decouple or be created {\textit{after}} 
the curvaton has decayed.

Conclusion 2, is particularly interesting from a particle physics point of view:

Firstly, it allows us to set an absolute upper bound for the creation temperature of dark matter, which has to be 
lower than $1.7 \times 10^6$ GeV for the case where the inflationary potential is quadratic. 
Gravitationally produced wimpzillas are in clear conflict with this bound and the assumptions of the model.
However, due to the strong model dependence of non-thermal dark matter 
generation, it is not straightforward to rule out any other class of such species. 

Secondly, it implies a relation between the temperature of creation of a cold dark matter candidate and the scale 
of inflation:
 \be
T_{\rm cdm}<1.6\times 10^4 \planck^{-\frac{3}{2}} H_*^{\frac{5}{2}}\cdot \left(\frac{100}{f_{\rm NL}}\right)
\ee
which links the physics of the early universe to the physics of the dark matter sector. This relation is shown to 
be robust against variations of the non-gaussian component, and would be specially significant if a
WIMP was observed in future experiments, since a portion of the space of parameters for this model 
would be ruled out. 

We remark the importance of the conclusions drawn above, because if 
the curvaton model turns out to be relevant inside the inflationary picture, this connection 
to dark matter physics could be the only probe into the scale of inflation. 

\section*{Acknowledgements}
I am very grateful to Wayne Hu and Rocky Kolb for very enlightening discussions and suggestions on this work.  
I would also like to thank to Juan Garc\'\i a-Bellido, Cora Dvorkin  and Amol Upadhye for useful comments on earlier 
versions of the draft. This work was supported by the Department of Energy.

\end{document}